\documentclass[aps,floatfix,showpacs,superscriptaddress,twocolumn]{revtex4} 
\usepackage{latexsym,graphicx,epsfig,here}

\usepackage{color}

\definecolor{darkgreen}{rgb}{0,0.5,0}
\definecolor{purple}{rgb}{0.5,0,0.5}
\definecolor{nblue}{rgb}{0.0,0.0,0.50}
\definecolor{scarlet}{rgb}{1.0,0.2,0}

\usepackage[colorlinks=true, pdfstartview=FitV, linkcolor=purple, citecolor= purple, urlcolor=blue]{hyperref}

\begin{document}

\title{B- to light-meson transition form factors}

\author{Mikhail A. Ivanov}
\affiliation{Bogoliubov Laboratory of Theoretical Physics, \\
Joint Institute for Nuclear Research, 141980 Dubna, Russia}
\author{J\"{u}rgen G. K\"{o}rner}
\affiliation{Institut f\"{u}r Physik, Johannes Gutenberg-Universit\"{a}t, \\
D-55099 Mainz, Germany}
\author{Sergey G. Kovalenko}
\affiliation{Centro de Estudios Subat\'omicos (CES),
Universidad T\'ecnica Federico Santa Mar\'\i a, \\
Casilla 110-V, Valpara\'\i so, Chile}
\author{Craig D. Roberts}
\affiliation{Physics Division, Argonne National Laboratory,
 Argonne IL 60439-4843, USA}
 
\begin{abstract}
We report a comprehensive set of results for $B$-meson heavy-to-light transition form factors calculated using a truncation of, and expression for, the transition amplitudes in which all elements are motivated by the study of Dyson-Schwinger equations in QCD.  In this relativistic approach, which realises confinement and dynamical chiral symmetry breaking, all physical values of momentum transfer in the transition form factors are simultaneously accessible.  Our results can be useful in the analysis and correlation of the large body of data being accumulated at extant facilities, and thereby in probing the Standard Model and beyond.
\end{abstract}

\maketitle

\section{Introduction}
\label{sec:intro}
Transition form factors that characterise the decays of $B$-mesons into light pseudoscalar and vector mesons, the so-called heavy-to-light decays, are basic to an understanding of this heavy-meson's exclusive semi-leptonic and rare radiative decays.  These form factors also provide the factorisable amplitudes that appear in $B$-meson exclusive nonleptonic charmless decays.  An understanding of all these processes is essential to the reliable determination of CKM matrix elements, and transitions mediated by electroweak and gluonic penguin operators.  Moreover, they should provide a means of searching for non Standard Model effects and $CP$ violation.  Considering all these factors, it is not surprising that heavy-light form factors are the subject of much experimental and theoretical scrutiny, as evidenced by the discussion in Ref.\,\cite{PDG}.

The analysis of heavy-to-light processes has two facets.  One is factorisation; viz., the feature that in exclusive decays of $B$-mesons there exist strong interaction effects that do not correspond to form factors.  These may be radiative corrections to purely hadronic operators in the weak effective Lagrangian or final state interactions between daughter hadrons \cite{El-Bennich:2006yi}.  The development of soft collinear effective-field theory (SCET) is providing a means of simplifying that problem, yielding factorisation theorems which enable a systematic approximation to be developed for a given process in terms of products of soft and hard matrix elements \cite{Feldmann:2006sf}.  Analyses relevant to the processes we consider herein may be found in Refs.\,\cite{SCET2,SCET1}; e.g., $B\to V\gamma$ decay amplitudes can be expressed in terms of a $B\to V$ form factor evaluated at the maximum recoil point, light-front distribution amplitudes of the heavy- and light-mesons and hard scattering kernels that can be evaluated perturbatively. 

The second facet, once factorisation for a given process is assumed or proved, is to evaluate the hadronic transition form factors.  Naturally, they cannot be calculated in perturbation theory.  The relevant matrix elements involve single hadrons in the initial and final states.  Hence, their calculation requires information about the structure of both heavy- and light-mesons.  A variety of theoretical approaches have been applied to this problem, recent amongst which are analyses using light-cone sum rules \cite{Khodjamirian:2006st,Ball:2004ye}, light-front quark models \cite{Lu:2007sg}, a constituent-quark model in a dispersion relation formulation \cite{Melikhov:2001zv}, and relativistic quark models -- e.g., Refs.\,\cite{Ebert:2006nz,Faessler:2002ut,Bc1,Bc2,Ivanov:2006ni}.  It is notable that while the methods of Refs.\,\cite{Khodjamirian:2006st,Ball:2004ye,Lu:2007sg} can only provide access to the form factors on a domain of small timelike $q^2$, the entire range of physical momenta is directly accessible in Refs.\,\cite{Melikhov:2001zv,Ebert:2006nz,Faessler:2002ut,Bc1,Bc2,Ivanov:2006ni}.  The latter is also true of the method employed in this article.

In the present context it is worth explaining that the relativistic constituent quark model introduced in Ref.\,\cite{RCQM} has been applied to the description of $B$ and $B_c$ transition form factors \cite{Faessler:2002ut,Bc1,Bc2,Ivanov:2006ni} using a small set of variable parameters.  The model's starting point is an interaction Lagrangian that describes the correlation of constituent-quarks within a meson and represents the system by a bound-state amplitude.  The so-called compositeness condition \cite{Z=0,Z=02} plays a key role in the consistent formulation of the model.  

In these studies the propagation of constituent-quarks is described by a free-particle Green function; i.e., 
\begin{equation}
S(k) = \frac{1}{\not\! k - m_Q},
\label{mishaprop}
\end{equation} 
wherein $m_Q$ is a light or heavy constituent-quark mass.  In order to avoid unphysical thresholds in transition amplitudes, it is necessary that for a meson of mass $m_H$ composed of quarks $Q_1$ and $Q_2$, $m_H<m_{Q_1}+m_{Q_2}$.  This poses problems for a description of light vector mesons ($\rho$, $K^\ast$), heavy flavoured vector mesons ($D^\ast$,$B^\ast$) and for $P$-wave and excited charmonium states.  To sidestep this, in the evaluation of matrix elements Refs.\,\cite{Bc1,Bc2} employed identical masses for all heavy pseudoscalar and vector flavored mesons; viz., $m_{B^\ast}=m_{B}$, $m_{D^\ast}=m_{D}$, and for all $P$-wave and excited charmonium states.  This is probably a reliable approximation for the heavy mesons because the corresponding mass splittings are small.  However, it is merely a stopgap measure for the light vector mesons, and one of the motivations for this article is to remedy that situation.  We implement confinement of light-quarks, in a manner which we shall subsequently elucidate.

Models based on results obtained via QCD's Dyson-Schwinger equations (DSEs) have also been employed \cite{misha1,misha2,mishasvy}.  These studies possess the feature that quark propagation is described by fully dressed Schwinger functions.  That dressing has a material impact on light-quark characteristics 
and, e.g., eliminates the threshold problem just described in connection with 
Refs.\,\cite{Faessler:2002ut,Bc1,Bc2,Ivanov:2006ni}.  Within this framework, as we have shown \cite{misha1,misha2,mishasvy} and shall see again herein, Eq.\,(\ref{mishaprop}) can nevertheless be justified for $b$-quarks and to some extent also for $c$-quarks.  Our purpose herein is to reappraise Ref.\,\cite{mishasvy} and extend that study to cover a fuller range of rare exclusive decays.  (NB. An introduction to DSEs can be found in Ref.\,\cite{cdragw} and their application in QCD is reviewed in Refs.\,\cite{bastirev,reinhardrev,pieterrev,christianrev,schladming}.)  

In Sec.\,\ref{sec:transitions} we describe the matrix elements that are the primary subject of this article, and introduce the approximation in which they are calculated.  The matrix elements are expressed in terms of dressed-quark propagators, meson Bethe-Salpeter amplitudes and interaction vertices.  These elements are explicated in Sec.\,\ref{sec:fit}.  We use forms determined and motivated by contemporary DSE studies.  Our calculated results are reported and explained in Sec.\,\ref{sec:calc}.  We wrap up in Sec.\,\ref{sec:end}.

\section{Heavy to light transitions}
\label{sec:transitions}
Herein our primary subjects are the following matrix elements, which can be expressed in Minkowski space via dimensionless form factors:
\begin{eqnarray}
\nonumber \lefteqn{\langle P(p_2)\,|\,\bar f_l\, \gamma_\mu\, b\,| B(p_1)\rangle}\\
& =& F_+(q^2) P_\mu+F_-(q^2) q_\mu\,, \label{ff}\\
\nonumber \lefteqn{
\langle P(p_2)\,|\,\bar f_l \,\sigma_{\mu\alpha}q^\alpha\, b\,|\,B(p_1)\rangle}\\
& =& 
\frac{i}{m_1+m_2}\left\{ q^2\, P_\mu - q^\nu P_\nu \, q_\mu\right\}\,F_T(q^2)\,,
\label{fT}\\
\nonumber 
\lefteqn{\langle V(p_2,\epsilon_2)\,|\,\bar f_l\, \gamma_\mu(1-\gamma_5)\, b\,|\,B(p_1)\rangle }\\
\nonumber &=& \frac{i}{m_1+m_2}\,\epsilon_2^{\dagger\nu}
\left\{-g_{\mu\nu}\,Pq\,A_0(q^2)+P_\mu P_\nu\,A_+(q^2) \right.\\
&& \left.
+q_\mu P_\nu\,A_-(q^2)+ i\varepsilon_{\mu\nu\alpha\beta} P^\alpha q^\beta  \,V(q^2)\right\}\,,\label{BVV}\\
\nonumber \lefteqn{
\langle V(p_2,\epsilon_2)\,|\,
\bar f_l\, \sigma_{\mu\nu}q^\nu(1+\gamma_5)\, b \,|\,B(p_1)\rangle}\\
\nonumber & =& \epsilon_2^{\dagger\nu}
\left\{-\left(g_{\mu\nu}-q_\mu q_\nu/q^2\right)\,q^\nu P_\nu \,a_0(q^2) \right.\\
\nonumber & &  +\left(P_\mu -q_\mu\,q^\rho P_\rho/q^2\right)\, P_\nu\,a_+(q^2) \\
&& \left.  +i\varepsilon_{\mu\nu\alpha\beta} P^\alpha q^\beta \,g(q^2)\right\}\,.
\label{BV}
\end{eqnarray}
(NB.\ The form factors defined in Eq.\,(\ref{BV}) satisfy the physical requirement $a_0(0)=a_+(0)$, which ensures that no kinematic singularity appears in the matrix element at $q^2=0$.  Constraint-free form factors for this transition were defined in Ref.\,\cite{Khodjamirian:2006st,Lu:2007sg}.)  In Eqs.\,(\ref{ff}) -- (\ref{BV}) one has a $B$-meson initial state with momentum $p_1$ and the symbols $P$, $V$ represent a light pseudoscalar or vector meson in the final state, $\bar f_l=\bar u,\bar d,\bar s$, with momentum $p_2$.  We define $P=p_1+p_2$, $q=p_1-p_2$, and denote by $\epsilon_2$ the polarisation four-vector of the vector meson.  Naturally, in Minkowski space the mass-shell conditions are $p_1^2=m_1^2$, $p_2^2=m_2^2$,  and $\epsilon_2^{\nu}p_{2\nu}=0$.  

For reference it is useful to relate the form factors we have defined to those used, e.g., in Ref.\,\cite{Khodjamirian:2006st}, which are denoted by a superscript $c$ in the following formulae:
\begin{eqnarray}
F_+ &=& f_+^{c}\,,\; 
%
F_T = f_T^{c}\,, \nonumber\\
\nonumber A_0 &=& \frac{m_1 + m_2}{m_1 - m_2}\,A_1^{c}\,, \; A_+ = A_2^{c}\,,\\
\nonumber A_- &=& \frac{2m_2(m_1+m_2)}{q^2}\,(A_3^{c} - A_0^{c})\,,\\
V &=& V^{c}\,, \nonumber\\
a_0 &=& T_2^{c}\,, \; g = T_1^{c}\,, \nonumber\\
a_+ &=& T_2^{c}+\frac{q^2}{m_1^2-m_2^2}\,T_3^{c}\,.
\end{eqnarray}
We note in addition that the form factors $A_i^c(q^2)$ satisfy
the constraints: $A_0^{c}(0)=A_3^{c}(0)$ and 
\begin{equation}
2m_2A_3^{c}(q^2) = (m_1+m_2) A_1^{c}(q^2) -(m_1-m_2) A_2^{c}(q^2)\,.
\end{equation}

\begin{figure}[t]
\centerline{\includegraphics[width=0.40\textwidth]{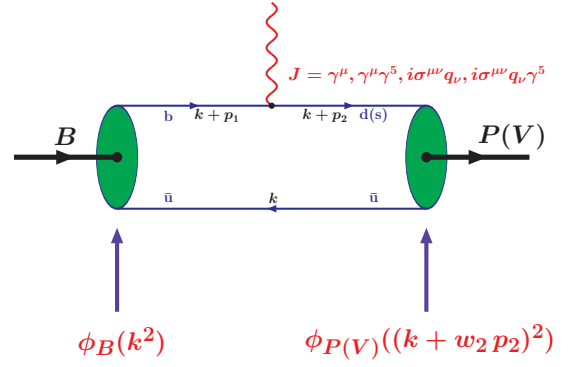}}

\caption{\label{Fig1} Diagrammatic representation of the matrix elements in Eqs.\,(\protect\ref{ff}) -- (\protect\ref{BV}).  In the figure, 
the solid lines denote dressed-quark propagators (Sec.\,\protect\ref{sec:Sp});
the filled ellipses, meson Bethe-Salpeter amplitudes (Sec.\,\protect\ref{sec:bsa}); and the connection of the undulating line with the dressed-quark propagator, an interaction vertex (Sec.\,\protect\ref{sec:vtx}).}
\end{figure}

The leading term in a systematic and symmetry preserving truncation of the DSEs yields a generalised impulse approximation to the matrix elements expressed in Eqs.\,(\ref{ff}) -- (\ref{BV}), which is depicted in Fig.\,\ref{Fig1}.  This diagram represents an amplitude via a single integral; i.e., \cite{fnEuclidean}
\newpage
\begin{eqnarray}
\nonumber
\lefteqn{{\cal A}(p_1,p_2) = {\rm tr}_{CD}\int\frac{d^4k}{(2\pi)^4}\,  \bar \Gamma_{P(V)}(k; -p_2) } \\
\nonumber &\times &  S_{f_l}(k+p_2)\Gamma_I(p_2,p_1) S_{b}(k+p_1)\Gamma_{B}(k;p_1) S_u(k)\,, \label{Amplitude}\\
\end{eqnarray}
where the trace is over colour and Dirac-spinor indices.  Equation~(\ref{Amplitude}) makes plain that our calculations require information about dressed-quark propagators -- $S(p)$, meson Bethe-Salpeter amplitudes -- $\Gamma(k;P)$, and interaction vertices -- $\Gamma_I(p,q)$.  In Sec.\,\ref{sec:fit} we discuss these in turn.

\begin{figure}[t]
\centerline{\includegraphics[width=0.45\textwidth]{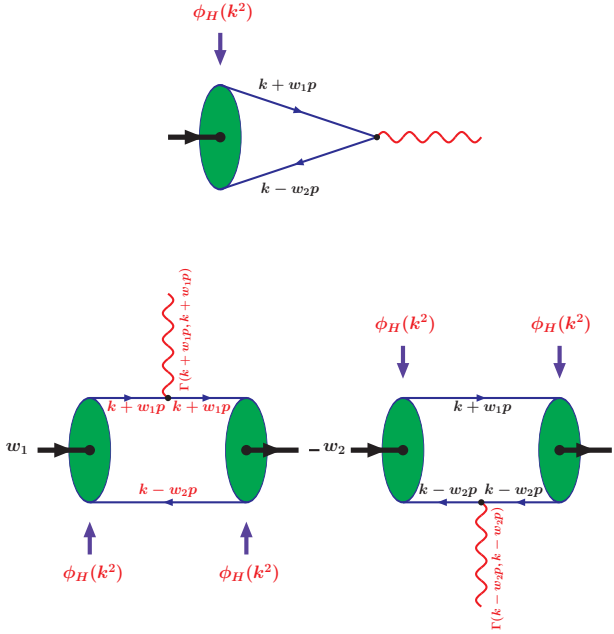}}

\caption{\label{Fig2} %
\emph{Upper diagram}: pictorial representation of a meson's leptonic decay constant, Eqs.\,(\protect\ref{f0m}) and (\protect\ref{fvm});
\emph{Lower diagram}: pictorial representation of the canonical normalisation condition for a meson's Bethe-Salpeter amplitude that is consistent with the generalised impulse approximation \protect\cite{cdrpion}, Eqs.\,(\protect\ref{norm1}) \& (\protect\ref{norm2}).  The vertex is defined in Eq.\,(\protect\ref{bcvtx}).
We indicate explicitly in the diagrams our choice of momentum partitioning when calculating the normalisation and leptonic decays: $w_1=1-w_2$, Sec.\,\protect\ref{sec:caveat}.}
\end{figure}

In order to provide a well-constrained analysis of the matrix elements in Eqs.\,(\ref{ff}) -- (\ref{BV}), we calculate simultaneously the leptonic decay constants of all participating mesons.  This decay proceeds as depicted in the upper diagram of Fig.\,\ref{Fig2}.  For pseudoscalar mesons that diagram represents \cite{fnrenormalisation}
\begin{equation}
\label{f0m}
 P_\mu f_{0^-}= \sqrt{2} {\rm tr}_{CD}\int\frac{d^4k}{(2\pi)^4}\, \gamma_5\gamma_\mu \chi^{0^-}(k;P)\,,
\end{equation}
with $\chi^H(k;P)= S_{f_1}(k+w_1 P) \Gamma^{H}(k;P) S_{f_2}(k - w_2 P)$ and $w_1+w_2 = 1$, whereas for vector mesons, 
\begin{equation}
\label{fvm}
 M_{1^-} f_{1^-}= \frac{\sqrt{2}}{3}{\rm tr}_{CD}\int\frac{d^4k}{(2\pi)^4}\, \gamma_\mu \chi_\mu^{1^-}(k;P) \,.
\end{equation}

For the Bethe-Salpeter amplitudes appearing in these expressions, the canonical normalisation condition consistent with Eqs.\,(\ref{ff}) -- (\ref{BV}) is depicted in the lower diagram of Fig.\,\ref{Fig2}, which for pseudoscalar mesons represents the expression
\begin{eqnarray}
\label{norm1}
2 P_\mu &=& \left[\frac{\partial}{\partial K_\mu} \Pi(P,K)\right]_{K=P}^{P^2=-m_{0^-}^2}\\
\nonumber \Pi(P,K)& =& {\rm tr}_{CD}\int\frac{d^4k}{(2\pi)^4}\,\bar \Gamma_{0^-}(k;-P)\\
\nonumber && \times  S_{f_1}(k+w_1 K) \Gamma_{0^-}(k;P) S_{f_2}(k-w_2 K)\,.\\ \label{norm2}
\end{eqnarray}
The expression for vector mesons is analogous and is given explicitly in Ref.\,\cite{mishasvy}.

\section{Propagators, amplitudes and vertices}
\label{sec:fit}
In this section we explain the dressed-quark propagators, the meson Bethe-Salpeter amplitudes, and the dressed-vertices describing the interaction between quarks and probes.

\subsection{Dressed-quark propagator} 
\label{sec:Sp}
For any quark flavour, the dressed-quark propagator has the general form 
\begin{eqnarray} 
\nonumber S(p) & = & -i \gamma\cdot p\, \sigma_V(p^2) + \sigma_S(p^2) \\
& = & 1/[i\gamma\cdot p\, A(p^2) + B(p^2)]. \label{SpAB} 
\end{eqnarray}
It is noteworthy that the mass function, $M(p^2)=B(p^2)/A(p^2)$, is independent of the renormalisation point.  

The propagator can be obtained from QCD's gap equation; namely, the DSE for the dressed-fermion self-energy, which for a given quark flavour in QCD is expressed as \begin{eqnarray}
S(p)^{-1} & =&  Z_2 \,(i\gamma\cdot p + m^{\rm bm}) + \Sigma(p)\,, \label{gendse} \\
\nonumber 
\Sigma(p) & = & Z_1 \int^\Lambda_q\! g^2 D_{\mu\nu}(p-q) \frac{\lambda^a}{2}\gamma_\mu S(q) \Gamma^a_\nu(q,p) , \\ 
&& \label{gensigma}
\end{eqnarray}
where $\int^\Lambda_q$ represents a Poincar\'e invariant regularisation of the integral, with $\Lambda$ the regularisation mass-scale \cite{mr98,mr97}, $D_{\mu\nu}(k)$ is the dressed-gluon propagator, $\Gamma_\nu(q,p)$ is the dressed-quark-gluon vertex, and $m^{\rm bm}$ is the quark's $\Lambda$-dependent bare current-mass.  The quark-gluon-vertex and quark wave function renormalisation constants, $Z_{1,2}(\zeta^2,\Lambda^2)$, depend on the renormalisation point, $\zeta$, the regularisation mass-scale and the gauge parameter.  The gap equation, Eq.\,(\ref{gendse}), is completely defined with the implementation of a renormalisation condition
\begin{equation}
\label{renormS} \left.S(p)^{-1}\right|_{p^2=\zeta^2} = i\gamma\cdot p +
m(\zeta)\,,
\end{equation}
where $m(\zeta)$ is the renormalised (running) mass: 
\begin{equation}
Z_2(\zeta^2,\Lambda^2) \, m^{\rm bm}(\Lambda) = Z_4(\zeta^2,\Lambda^2) \, m(\zeta)\,,
\end{equation}
with $Z_4$ the Lagrangian-mass renormalisation constant.  It is important that in QCD the chiral limit is strictly and unambiguously defined by \cite{mr98,mr97}
\begin{equation}
\label{limchiral}
Z_2(\zeta^2,\Lambda^2) \, m^{\rm bm}(\Lambda) \equiv 0 \,, \forall \Lambda \gg \zeta \,,
\end{equation}
which states that $\hat m = 0$, where $\hat m$ is the renormalisation-point-invariant current-quark mass.

The gap equation has been much studied and the features of its solution elucidated.  Hereafter, we explain the manner in which phenomenological studies can capitalise on this.

\subsubsection{Light-quarks}
It is a longstanding prediction of DSE studies in QCD that for light-quarks the wave function renormalisation and dressed-quark mass: 
\begin{equation} 
\label{ZMdef}
Z(p^2)=1/A(p^2)\,,\;M(p^2)=B(p^2)/A(p^2)\,, 
\end{equation} 
respectively, receive strong momentum-dependent corrections at infrared momenta \cite{lane,politzer,cdragw}: $Z(p^2)$ is suppressed and $M(p^2)$ enhanced.  These features are an expression of dynamical chiral symmetry breaking (DCSB) (and, plausibly, of confinement -- see below).  The enhancement of $M(p^2)$ is central to the appearance of a constituent-quark mass-scale and an existential prerequisite for Goldstone modes.  The mass function evolves with increasing $p^2$ to reproduce the asymptotic behaviour familiar from perturbative analyses, and that behaviour is unambiguously evident for $p^2 \gtrsim 10\,$GeV$^2$ \cite{mishasvy}. These DSE predictions are confirmed in numerical simulations of lattice-regularised QCD \cite{bowman}, and the conditions have been explored under which pointwise agreement between DSE results and lattice simulations may be obtained \cite{bhagwat,bhagwat2,alkoferdetmold,bhagwattandy}.

The impact of this infrared dressing on hadron phenomena has long been emphasised \cite{cdrpion} and, while numerical solutions of the quark DSE are now readily obtained, the utility of an algebraic form for $S(p)$ when calculations require the evaluation of numerous integrals is self-evident.  An efficacious parametrisation of $S(p)$, which exhibits the features described above, has been used extensively in hadron studies \cite{bastirev,reinhardrev,pieterrev,schladming}.  It was employed in Ref.\,\cite{mishasvy} and is expressed in an algebraic form via entire functions \cite{mark}:
\begin{eqnarray} 
\nonumber \bar\sigma_S(x) & =&  2\,\bar m \,{\cal F}(2 (x+\bar m^2)) \\
&& + {\cal F}(b_1 x) \,{\cal F}(b_3 x) \,  
\left[b_0 + b_2 {\cal F}(\epsilon x)\right]\,,\label{ssm} \\ 
\label{svm} \bar\sigma_V(x) & = & \frac{1}{x+\bar m^2}\, \left[ 1 - {\cal F}(2 (x+\bar m^2))\right]\,, 
\end{eqnarray}
with $x=p^2/\lambda^2$, $\bar m$ = $m/\lambda$, 
\begin{equation}
\label{defcalF}
{\cal F}(x)= \frac{1-\mbox{\rm e}^{-x}}{x}  \,, 
\end{equation}
$\bar\sigma_S(x) = \lambda\,\sigma_S(p^2)$ and $\bar\sigma_V(x) = \lambda^2\,\sigma_V(p^2)$.  The parameter values were fixed in Ref.\,\cite{mishasvy} by requiring a least-squares fit to a wide range of light- and heavy-meson observables, and take the values:
\begin{equation} 
\label{tableA} 
\begin{array}{llcccc} 
f &   \bar m_f& b_0^f & b_1^f & b_2^f & b_3^f \\\hline 
u=d &   0.00948 & 0.131 & 2.94 & 0.733 & 0.185 \\
s &   0.210   & 0.105 & 3.18 & 0.858 & 0.185 
\end{array}\;.
\end{equation}
The mass-scale $\lambda=0.566\,$GeV, with which value the current-quark masses are \begin{equation}
m_u=5.4\,{\rm MeV}, \; m_s=119\,{\rm MeV}.
\end{equation}
In addition one obtains the following Euclidean constituent-quark masses, defined as the solution of $(M_f^E)^2 = M_f(M_f^2)^2$:
\begin{equation}
\label{MQ}
M_u^E = 0.36\,{\rm GeV}\,,\; 
M_s^E = 0.49\,{\rm GeV}\,.
\end{equation}

Equations (\ref{svm}) -- (\ref{tableA}) provide an algebraic form for $S(p)$ that combines the effects of confinement and DCSB with free-particle behaviour at large spacelike $p^2$.  For instance, it is noteworthy that, motivated by DSE studies \cite{entire1,entire2}, Eqs.~(\ref{ssm}) and (\ref{svm}) express the dressed-quark propagator as an entire function.  Hence $S(p)$ does not have a Lehmann representation, which is a sufficient condition for confinement because of the associated violation of reflection positivity.  This notion may be traced from Refs.\,\cite{entire1,entire2,stingl,krein} and is reviewed in Refs.\,\cite{cdragw,bastirev,reinhardrev}.  Additional commentary on this point is provided by Refs.\,\cite{hawes1,hawes2}. 

As explained, e.g.\ in Ref.\,\cite{mishasvy}, one expression of DCSB in our parametrisation of $S(p)$ is an expression for the chiral limit vacuum quark condensate:
\begin{equation}
-\langle \bar u u \rangle_\zeta = \lambda^3 \frac{3}{4\pi^2} \frac{b_0^u}{b_1^u b_2^u}\ln\frac{\zeta^2}{\Lambda_{\rm QCD}^2}\,,
\end{equation}
which assumes the value ($\Lambda_{\rm QCD} = 0.2\,{\rm GeV}$) \cite{cdrpion}
\begin{equation}
-\langle \bar u u \rangle_{\zeta=1\,{\rm GeV}} = (0.22\,{\rm GeV})^3.
\end{equation}
A detailed discussion of the vacuum quark condensate in QCD can be found in Refs.\,\cite{langfeld,changlei}.

\subsubsection{Heavy-quarks}
While the impact of DCSB on light-quark propagators is marked, that is not true for heavier quarks, as can be seen by considering the dimensionless and renormalisation-group-invariant ratio $\varsigma_f:=\sigma_f/M^E_f$, where $\sigma_f$ is a constituent-quark $\sigma$-term \cite{wrightLC05}.  This ratio measures the effect of explicit chiral symmetry breaking on the dressed-quark mass-function compared with the sum of the effects of explicit and dynamical chiral symmetry breaking.  Naturally, $\varsigma_f$ must vanish for light-quarks because the magnitude of their constituent-mass owes primarily to dynamical chiral symmetry breaking.  For heavy-quarks, $\varsigma_f$ approaches one.  The ratio $\varsigma_f$, calculated using the DSE model introduced in Ref.\,\cite{Maris:1999nt}, was discussed in Ref.\,\cite{schladming}: $\varsigma_u = 0.02$, $\varsigma_s = 0.23$, $\varsigma_c = 0.65$, $\varsigma_b = 0.8$.  

It follows that for the $c$-quark it is reasonable and for the $b$-quark, sensible, to employ a constituent-quark propagator; viz.,
\begin{equation}
\label{SQ}
S_{Q}(k) = \frac{1}{i \gamma\cdot k + \hat M_Q}\,,\; Q=c,b,
\end{equation}
where the values
\begin{equation}
\hat M_c = 1.32\,{\rm GeV}, \; 
\hat M_b = 4.65\,{\rm GeV},
\end{equation}
were fixed in the same least-squares fit as the light-quark parameters in Eq.\,(\ref{tableA}) \cite{mishasvy}.

\subsection{Bethe-Salpeter amplitudes}
\label{sec:bsa}
The meson Bethe-Salpeter amplitudes, which appear in Fig.\,\ref{Fig1} and are consistent with the generalised impulse approximation, are properly obtained from the improved-ladder Bethe-Salpeter kernel, as described for example in Ref.\,\cite{mr97}.  The solution of this equation requires a simultaneous solution of the quark DSE.  However, since we have already chosen to simplify the calculations by parametrising $S(p)$, we follow Ref.~\cite{mishasvy} and also employ that expedient with $\Gamma_{P(V)}$.

\subsubsection{Light pseudoscalar mesons}
Dynamical chiral symmetry breaking and the axial-vector Ward-Takahashi identity have a big impact on the structure and properties of light pseudoscalar mesons.  In fact, the quark-level Goldberger-Treiman relations derived in Ref.\,\cite{mr98} motivate and support the following efficacious parametrisation of light pseudoscalar meson Bethe-Salpeter amplitudes:
\begin{eqnarray}
\label{piKamp}
\Gamma_{P}(k;P) &= &i\gamma_5\,{\cal E}_{P}(k^2)\,,\; P=\pi,K,\\
{\cal E}_{P}(k^2)  &= & \frac{\surd 2}{f_{P}}\,B_{P}(k^2)\,, \label{E=B}
\end{eqnarray}
where $B_P:=\left.B_u\right|_{b_0^u\to b_0^P}$ is 
obtained from Eqs.~(\ref{SpAB}), (\ref{ssm}), (\ref{svm}) via the replacements 
\begin{equation}
b_0^u \rightarrow b_0^{\pi}=0.204\;\; \mbox{and} \;\; b_0^u \rightarrow b_0^{K}=0.319\,, \label{b0u}
\end{equation} as appropriate.  We emphasise that Eq.\,(\ref{E=B}) expresses the intimate connection between the leading covariant in a pseudoscalar meson's Bethe-Salpeter amplitude and the scalar piece of the dressed-quark self energy.  As usual, the values of the parameters, Eq.\,(\ref{b0u}), were fixed in the same least-squares fit as the light-quark parameters in Eq.\,(\ref{tableA}) \cite{mishasvy}.

\subsubsection{Light vector mesons}
\label{lvm}
Dyson-Schwinger equation studies of the structure and properties of light-vector mesons are available in Refs.\,\cite{bhagwatmaris,Maris:1999nt,marisdq,jarecke}.  A consideration of these studies indicates that in connection with phenomena that are predominantly governed by infrared mass-scales, which is typically the case for form factors on the physical domain of accessible timelike momenta, it is reasonable to parametrise vector meson Bethe-Salpeter amplitudes as follows:
\begin{equation}
\label{gammaV}
\Gamma^V_{\mu}(k;p) = \frac{1}{{\cal N}_V}\,
\left(\gamma_\mu + p_\mu\,\frac{\gamma\cdot p}{M_V^2}\right)\,
\varphi(k^2)\,,
\end{equation}
with 
\begin{equation}
\label{gammaVnew}
\varphi(k^2) = \exp(-k^2/\omega_V^2)\,;
\end{equation}
namely, a function whose support is concentrated in the infrared.
Clearly, $p_\mu \Gamma^V_{\mu}(k;p) = 0$.    
Since such phenomena are at the heart of our study, we employ this expedient herein.  Hence we have an \textit{Ansatz} that is little different to that used in Ref.\,\cite{mishasvy}.  The one parameter is a mass-scale $\omega_V$, which specifies the momentum space width of the amplitude.  The normalisation is calculated via the analogues of Eqs.\,(\ref{norm1}) and (\ref{norm2}).

\subsubsection{Heavy mesons}
\label{sec:hm}
For heavy mesons we write \cite{mishasvy}
\begin{equation}
\label{HPamp}
\Gamma_H(k;p)= \frac{1}{{\cal N}_H}\,\gamma_H\,
\varphi_H(k^2)\,,
\end{equation}
where $\gamma_P=i\,\gamma_5$, $\gamma_V=\gamma\cdot\epsilon_V$, $\varphi_H(k^2) = \exp\left(-k^2/\omega_H^2\right)$ and ${\cal N}_H$ is fixed by Eqs.\,(\ref{norm1}) and (\ref{norm2}).  In common with Ref.\,\cite{mishasvy}, we assume that the width parameters are spin independent; i.e., $\omega_{B^\ast}=\omega_B$, $\omega_{D^\ast}=\omega_D$, as would be the case were heavy-quark symmetry to be realised exactly.  On the other hand, we allow full flavour-dependence, since high precision experiments related to heavy-quark systems are becoming available.  Thus we fit with $\omega_D$, $\omega_{D_s}$, $\omega_B$, $\omega_{B_s}$ treated as independent parameters. 

\subsubsection{Caveat: momentum partitioning}
\label{sec:caveat}
Manifest Poincar\'e covariance is a feature of the direct application of DSEs to the calculation of hadron observables.  This is illustrated, e.g., in Refs.\,\cite{mr97,oettelB,maristandyLC,bhagwatmaris}, which also emphasise that manifest covariance is only possible if the complete and complicated structure of hadron bound-state amplitudes is retained.  That can impose numerical costs since, e.g., the complete vector meson Bethe-Salpeter amplitude involves eight Poincar\'e covariants.

Herein, we use simple one-covariant models for the amplitudes with a goal of describing simultaneously a wide range of phenomena.  With the omission of the full structure of amplitudes, however, comes the complication that our results can be sensitive to the definition of the relativistic relative momentum.  \emph{Every} study that fails to retain the full structure of the Bethe-Salpeter amplitude shares this complication. 

Hence to proceed we must specify the relative momentum.  In common with Refs.\,\cite{mishasvy,misha1,misha2}, when a heavy-quark line is involved, we allocate all the heavy-light-meson's momentum to that heavy-quark and choose the single covariant in the heavy-light-meson's Bethe-Salpeter amplitude to be a function of only the light-quark momentum.  This is evident in Fig.\,\ref{Fig1}.  

For the light mesons, our choice is clearly depicted in Fig.\,\ref{Fig2}; viz., a quark with momentum $k_1$ and antiquark with momentum $k_2$ are bound into a system with total momentum $p$, with the relative momentum $k=w_2 k_1 + w_1 k_2$, where $w_1 + w_2=1$.  In Refs.\,\cite{misha1,misha2,mishasvy}, and in other phenomenological DSE studies, e.g., \cite{cdrpion,mark,pionvalence}, $w_2=1/2$ is usually implicit.  Herein we allow $w_2$ to vary and do not require $w_2^{ud} = w_2^{us}$.  

We re-emphasise that the existence of optimal values for these parameters is a consequence of the truncations employed in setting up the bound state model.  No dependence would exist in an \textit{ab initio} study, which is possible, e.g., Ref.\,\cite{bhagwatmaris,mindgap}.  However, for the breadth of application herein, an \textit{ab initio} study is beyond \textit{our} capacity.

\subsection{Dressed vector vertices}
\label{sec:vtx}
The dressed-quark-photon vertex has been much studied, with direct numerical solutions of the relevant inhomogeneous Bethe-Salpeter equation providing valuable information and delivering reliable predictions of meson form factors \cite{bhagwatmaris,marisG1,marisG2}.  However, since we have parametrised the dressed-quark propagators we follow Ref.\,\cite{cdrpion} and employ the Ball-Chiu \textit{Ansatz} \cite{bc80}: 
\begin{eqnarray}
\nonumber \lefteqn{i\Gamma^f_\nu(\ell_1,\ell_2) = i\Sigma_A(\ell_1^2,\ell_2^2)\,\gamma_\mu}\\
\nonumber
&& +
(\ell_1+\ell_2)_\mu\,\left[\frac12 i\gamma\cdot (\ell_1+\ell_2) \,
\Delta_A(\ell_1^2,\ell_2^2) + \Delta_B(\ell_1^2,\ell_2^2)\right],\\
&& \label{bcvtx}
\end{eqnarray}
where 
\begin{eqnarray}
\Sigma_F(\ell_1^2,\ell_2^2)& = &\frac12\,[F(\ell_1^2)+F(\ell_2^2)]\,,\;\\
\Delta_F(\ell_1^2,\ell_2^2) & = &
\frac{F(\ell_1^2)-F(\ell_2^2)}{\ell_1^2-\ell_2^2}\,,
\end{eqnarray}
with $F= A_f, B_f$; i.e., the scalar functions in Eq.~(\ref{SpAB}) evaluated with the appropriate dressed-quark propagator.   It is critical that this \textit{Ansatz} satisfies the Ward-Takahashi identity, 
\begin{equation}
\label{vwti}
(\ell_1 - \ell_2)_\nu \, i\Gamma^f_\nu(\ell_1,\ell_2) = 
S_f^{-1}(\ell_1) - S_f^{-1}(\ell_2)\,,
\end{equation}
which is a sufficient condition to guarantee current conservation \cite{cdrpion}, and very useful in that it is completely determined by the dressed-quark propagators.  The \textit{Ansatz} has been used fruitfully in many hadronic applications; e.g., \cite{cdrpion,mark,pionvalence}.  

It is noteworthy that for heavy-quarks, since from Eq.\,(\ref{SQ}) one has $A_Q(k^2)\equiv 1$ and $M_Q(k^2)\equiv \hat M_Q$, Eq.\,(\ref{bcvtx}) yields
\begin{equation}
\Gamma^Q_\mu(\ell_1,\ell_2) = \gamma_\mu\,;
\end{equation}
namely, the correct heavy-quark limit.  

We capitalise on this throughout; namely, we replace dressed-vertices on heavy-quark lines by their bare form.

%

\begin{table}[t]
\caption{Leptonic decay constants $f_H$ (MeV) calculated using the parameter values listed in connection with Eq.\,(\ref{params}).  Data and selected calculations are provided for comparison.  The compilation of charm meson results in Ref.\,\cite{Zweber:2007dy} was useful in our analysis.  
}
\label{tab:leptonic}
\begin{center}
\def\arraystretch{1.1}
\begin{tabular}{|c|c|ll|}
\hline
    & This work  & \hspace*{1cm} Other & Reference  \\
\hline
 $f_\rho$     & 209 & 209(2) & PDG \cite{PDG} \\
 $f_{K^\ast}$ & 217 & 217(5) & PDG \cite{PDG} \\
\hline
$f_D$
 & 223 & $222.6\,(16.7)^{+2.8}_{-3.4}$  & CLEO \cite{CLEOfDs,CLEOStoneICHEP,CLEOfD}  \\
 &     & $371^{+129}_{-117}(25)$      & BES \cite{BESfD}       \\
 &     & 227 &   RCQM \cite{Ivanov:2006ni} \\\hline
$f_{D_s}$
 & 281 &  280(12)(6)      & CLEO \cite{CLEOfDs,CLEOStoneICHEP,CLEOfD}  \\
 &     & 283(17)(4)(14)   & BaBar \cite{BABARfDs} \\ 
 &     & 255 &   RCQM \cite{Ivanov:2006ni} \\\hline
$\displaystyle\frac{f_{D_s}}{f_D}$ 
 & 1.26 & 1.26(11)(3) &  CLEO \cite{CLEOfDs,CLEOStoneICHEP,CLEOfD}  \\
 &      & 1.12        &  RCQM \cite{Ivanov:2006ni}                  \\\hline
$f_{D^\ast}$
 & 321 & $245\,(20)^{+3}_{-2}$ & LAT \cite{Becirevic:1998ua}\\
 &     &   249                  &  RCQM \cite{Ivanov:2006ni}  \\\hline
$f_{D^\ast_s}$
 & 364 & $272\,(16)^{+3}_{-20}$   &  LAT \cite{Aubin:2005ar} \\
 &     &  266                   &  RCQM \cite{Ivanov:2006ni}  \\\hline
$f_B$
 & 176  & $229^{+36}_{-31}\,({\rm stat})^{+34}_{-37}\,({\rm syst})$ & 
BELLE\cite{Ikado:2006un} \\  
 &      & $216(9)(19)(4)(6)$ & HPQCD LAT \cite{Gray:2005ad} \\
 &      & $177\,(17)^{+22}_{-22}$    & UKQCD LAT \cite{Lellouch:2000tw}\\
 &      & $179 \,(18)^{+34}_{-9}$     & LAT \cite{Becirevic:1998ua}\\
 &      &  $210.5(11.4)(5.7)$         & Chiral Lat\cite{Guo:2006nt} \\
 &      &  187                       & RCQM \cite{Ivanov:2006ni}  \\\hline
$f_{B_s}$
 & 211 & $259\,(32)$                 & HPQCD LAT \cite{Gray:2005ad} \\
 &     & $260 \,(7)\,(26)\, (8)\,(5)$ & LAT \cite{Wingate:2003gm} \\
 &     & $204 \,(12)^{+24}_{-23}$     & UKQCD LAT \cite{Lellouch:2000tw}\\
 &     & $204 \,(16)^{+36}_{-0}$      & LAT \cite{Becirevic:1998ua}\\
 &      &  218                         & RCQM \cite{Ivanov:2006ni}  \\\hline
$\displaystyle\frac{f_{B_s}}{f_B}$
 & 1.20 & $ 1.20\,(0.03)\,(0.01) $        & HPQCD LAT \cite{Gray:2005ad} \\
 &      & $1.15 \,(0.02)^{+0.04}_{-0.02}$  & UKQCD LAT \cite{Lellouch:2000tw}\\
 &      & $1.14 \,(0.03)^{+0.01}_{-0.01}$  & LAT \cite{Becirevic:1998ua}\\
 &      &   1.16                         & RCQM \cite{Ivanov:2006ni}  \\\hline
$f_{B^\ast}$
 & 198  & $196 \,(24)^{+39}_{-2}$    & LAT \cite{Becirevic:1998ua}\\
 &      &  196                       & RCQM \cite{Ivanov:2006ni}  \\\hline
$f_{B^\ast_s}$
 & 235  & $229 \,( 20)^{+41}_{-16}$  & LAT \cite{Becirevic:1998ua}\\
 &      &  229                       & RCQM \cite{Ivanov:2006ni}  \\
\hline
\end{tabular}
\end{center}
\end{table}

\subsection{Heavy-quark symmetry limits}
With algebraic parametrisations of each of the pieces that comprise a matrix element one can obtain simple formulae that express the heavy-quark symmetry limits of these matrix elements.  References~\cite{misha1,misha2,mishasvy} detail the results of such analysis.  In particular, Sec.\,VI of Ref.\,\cite{mishasvy} provides a complete discussion of the heavy-quark symmetry limits of numerous matrix elements.  Moreover, Sec.\,III therein describes a novel result for pseudoscalar meson masses in the heavy-quark limit, first described in Ref.\,\cite{marisrobertsheavy}.

To highlight a couple of results relevant to our present discussion, we observe that the leptonic and semileptonic decays of heavy mesons were considered in Ref.\,\cite{misha1}.  In accord with heavy-quark effective theory \cite{Neubert},  it was shown \cite{misha1} that in the heavy-quark limit the leptonic decay constants evolve as $(\hat M_Q)^{-1/2}$ and the matrix elements describing semileptonic heavy-heavy decays can be expressed in terms of a single universal function, $\xi$.  The calculated result for this function can be written 
\begin{eqnarray}
\nonumber \xi(w)  &= & \frac{N_c}{4\pi^2}\,\kappa_1 \kappa_2
\int_0^1 d\tau\,\frac{1}{W}\,
\int_0^\infty du \, \varphi_H^2(z_W)\\\
&& \times \left[\sigma_S(z_W) + \sqrt{\frac{u}{W}} \sigma_V(z_W)\right]\,,
\label{xif}
\end{eqnarray}
with $\kappa^2=1/[m_{H} {\cal N}_{H}^2]$, $W= 1 + 2 \tau (1-\tau)
(w-1)$, $z_W= u - 2 E_H \sqrt{u/W}$, and $w= - v_{H_1} \cdot v_{H_2}$ where $v_H=p_H/m_H$.  Owing to the canonical normalisation of the Bethe-Salpeter amplitudes, $\xi(w=1)=1$.  

A determination of this function based on the $B\to D$ transition yields \cite{mishasvy} a numerical result that is accurately interpolated by
\begin{equation}
\xi(w)= \frac{1}{1+ \rho^2 (w-1)},\; \rho^2=1.98\,,
\end{equation}
with here $w= (m_B^2+m_D^2-t)/(2 m_B m_D)$.  A comparison of this result with inferences from experiment is presented in Refs.\,\cite{misha2,mishasvy}.  

In addition to reproducing the results of heavy-quark symmetry, it is an important feature of the DSEs that one can examine the fidelity of the formulae obtained in the heavy-quark limit; viz., elucidate the extent to which they are physically realised.  Reference~\cite{mishasvy} provides a unified and uniformly accurate description of a broad range of light- and heavy-meson observables.  It concludes that corrections to the heavy-quark symmetry limit of $\lesssim 30$\% are encountered in $b\to c$ transitions and that these corrections can be as large as a factor of $2$ in $c\to d$ transitions. 

\section{Calculated Results}
\label{sec:calc}
\subsection{Parameters and fitting}

In the framework we have set up there are eight variable parameters: The widths of the light-vector-meson Bethe-Salpeter amplitudes -- $\omega_\rho$, $\omega_{K^\ast}$, Sec.\,\ref{lvm}; the widths of the heavy-meson Bethe-Salpeter amplitudes --  $\omega_D$, $\omega_{D_s}$, $\omega_{B}$, $\omega_{B_s}$, Sec.\,\ref{sec:hm}; and the light-quark momentum partitioning parameters -- $w_2^{\rm ud}$,  $w_2^{\rm us}$, Sec.\,\ref{sec:caveat}.  All other parameters, including the quark masses, are taken as reported in Ref.\,\cite{mishasvy}.  We observe that in the widespread application of this framework and variants thereupon, calculations, when they can be compared with observables, are accurate to a root-mean-square (rms) deviation of 15\%.  The current application is not different in principle and hence this value should provide a reasonable estimate of our theoretical error.

We determine our parameters through a least-squares fit to meson leptonic decay constants, expressions for which are given in Eqs.\,(\ref{f0m}) and (\ref{fvm}), and illustrated in Fig.\,\ref{Fig2}.  The parameter values obtained via this procedure are $w_2^{\rm ud}=0.377$,  $w_2^{\rm us}=0.316$ and, in GeV:
\begin{equation}
\begin{array}{c|c|c|c|c|c}
\omega_\rho & \omega_{K^\ast} & \omega_D & \omega_{D_s} & \omega_{B}  & \omega_{B_s} \\\hline
0.561 & 0.611 & 1.50 & 1.97 & 1.37 &  1.63
\end{array}\,.
\label{params}
\end{equation}
The ordering is of interest, as may be seen by defining a matter-radius scale: $\ell_H=1/\omega_H$.  Apparently, $\ell_{K^\ast} = 0.32\,{\rm fm} < \ell_\rho= 0.35\,{\rm fm}$, and the ratio of these two scales is in accordance with the ratio of charge radii reported in Ref.\,\cite{bhagwatmaris}.  Moreover, $\ell_{D_s} = 0.10\,{\rm fm} < \ell_D= 0.13\,{\rm fm}$ with a ratio $0.76$, and $\ell_{B_s} = 0.12\,{\rm fm} < \ell_B= 0.14\,{\rm fm}$ with a ratio $0.84$.  Hence, the ordering within $D$- and $B$-meson systems is consistent with intuition.  However, not so the result that systems containing a $c$-quark are, by this rudimentary measure, smaller than systems containing a $b$-quark.  We expect that a more sophisticated representation of heavy-quark propagators and heavy-meson Bethe-Salpeter amplitudes will reverse this aspect of the parametrisation.

The calculated values of observables are: $\rho\to\pi\pi$ coupling constant $g_{\rho\pi\pi}=4.62$ [expt.\ =$5.92(2)$] and $K^\ast\to K\pi$ coupling  $g_{K^\ast K\pi}=4.55$ [expt.\ =$4.67(4)$], expressions for which are provided in Ref.\,\cite{mishasvy}, plus the leptonic decay constants listed in Table~\ref{tab:leptonic}.  For ease of reference, we also list in Table~\ref{tab:leptonic} the leptonic decay constants calculated in Ref.\,\cite{Ivanov:2006ni}; i.e., the relativistic constituent-quark model described in Sec.\,\ref{sec:intro}.  NB.\ Our calculated light-meson leptonic decay constants are unchanged from Ref.\,\cite{mishasvy}: $f_\pi=146\,$MeV; $f_K=178\,$MeV --  cf.\ expt.: $131\,$MeV and $161\,$MeV, respectively.  

A comparison between our results in Table~\ref{tab:leptonic} and those taken from elsewhere yields $\chi^2/\mbox{\rm degree-of-freedom}\sim 12$ and $\chi^2/\mbox{\rm number-of-observables}\sim 4$.  It is notable that omitting those entries in the table for which experimental results are not available, we have 
$\chi^2/\mbox{\rm number-of-observables}= 0.2$.  Plainly, the lattice results are providing relevant constraints.  Hence, improved reliability of such studies would be welcome.

\begin{figure}[tb]
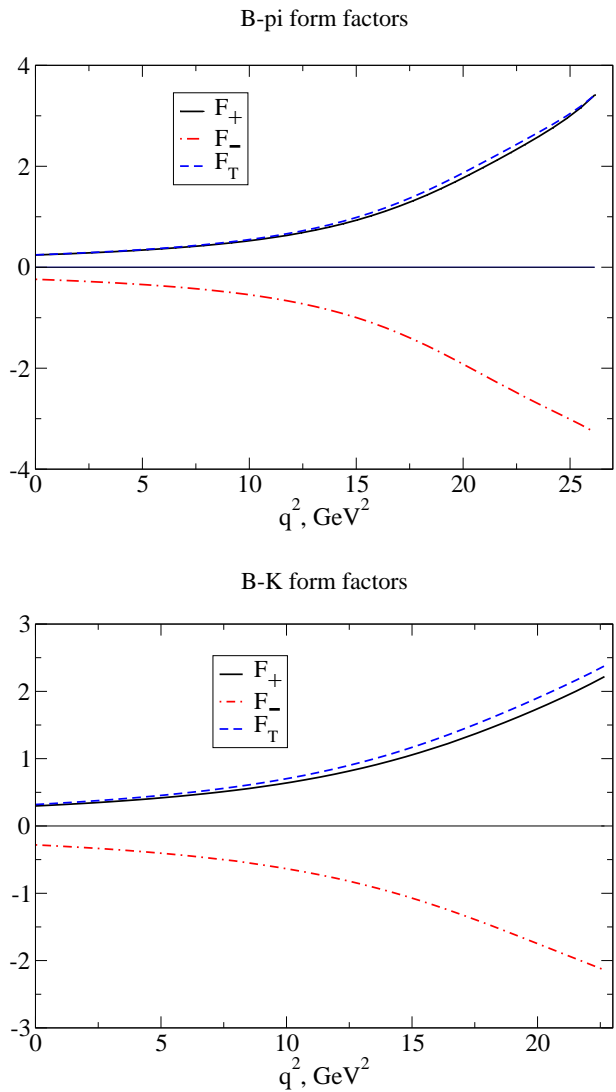

\begin{center}
\hspace*{-0.5cm}
\begin{tabular}{c}
\includegraphics[width=0.45\textwidth]{Fig3aBpi.eps}\\[3ex]

\includegraphics[width=0.45\textwidth]{Fig3bBK.eps}
\end{tabular}
\end{center}
\caption{\label{fig:formfactors1} Our results for the form factors appearing in Eqs.\,(\protect\ref{ff}) \& (\protect\ref{fT}) -- \emph{Top panel}, $B\to \pi$; and \emph{bottom panel}, $B\to K$: \emph{solid} $F_+$, \emph{dot-dashed} $F_-$, \emph{dashed} $F_T$. }
\end{figure}

\subsection{$B\to P(V)$ transitions}
\label{sec:formfac}
With all parameters fixed, we now proceed to the calculation of the $B\to P(V)$ heavy-to-light transitions depicted in Fig.\,\ref{Fig1}.  

\subsubsection{Technical remark} 
Before reporting the predictions, we offer a technical remark.  Calculation of the generalised impulse approximation to the transitions involves the numerical evaluation of a four-dimensional integral whose integrand is the convolution of entire functions and functions with a simple pole.  The straightforward use of spherical coordinates in the Euclidean loop integral and the choice of the $B$-meson rest-frame works well only for $p_1^2\le M^2_b$.  When $p_1^2> M^2_b$ one needs to shift the integration contour into the complex plane.  Since that is not easily done numerically, we employ an alternate representation that can be used straightforwardly for any $p_1^2$.  Namely, with $p_1^2=-m_1^2$, $p_2^2=-m_2^2$ and $(p_1-p_2)^2=-q^2$:  
\begin{eqnarray}
\nonumber\lefteqn{
\int\frac{d^4 k}{\pi^2} \frac{ F(k^2)\, \phi\Big((k+w_2 p_2)^2\Big) \sigma\Big((k+p_2)^2\Big)  }
{M^2_b+(k+p_1)^2}}\\
&&  =
\frac{1}{\pi}\int\limits_0^\infty du \int\limits_0^1 dv \int\limits_0^\pi d\theta
F(z_1) \, \phi(y_2) \, \sigma(z_2)\,, 
\end{eqnarray}
where the new variables are 
\begin{eqnarray}
z_1 &=& u+\frac{1-v}{v}[M^2_b-v m_1^2]\,,  \\
\nonumber y_2 &=& z_1+2 i \sqrt{u v} \cos\theta\,w_2 m_2 \\
&& +(1-v)\,w_2\, (m_1^2+m_2^2-q^2)-w_2^2\,m_2^2\,,\\
\nonumber z_2 &=& z_1+2 i \sqrt{u v} \cos\theta\,m_2 \\
&& +(1-v)\, (m_1^2+m_2^2-q^2)-m_2^2\,.
\end{eqnarray}

\begin{table}[t]
\caption{\label{tab:param_pi_rho} $B\to \pi(\rho)$ transition form factors: values of the parameters in the interpolating function, Eq.\,(\protect\ref{parametrization}).  Our results are marked by an asterisk.  For comparison, where available we also give analogous values inferred from the fit of Ref.\,\cite{Melikhov:2001zv}.}
\begin{center}
\begin{tabular}{lrcc|c}
    & $F(0)$   & $a$  & $b$ & source  \\ \hline
$F_+$ &  0.24  & 1.87 & 0.93  &  $\ast$  \\
      &  0.29  & 1.48 & 0.48  & \cite{Melikhov:2001zv} \\
$F_-$ & -0.24  & 1.97 & 1.04  & $\ast$ \\
$F_T$ &  0.24  & 1.92 & 1.00  & $\ast$ \\
      &  0.28  & 1.48 & 0.48  & \cite{Melikhov:2001zv}\\\hline
$A_0$ &  0.32  & 1.16 & 0.32  & $\ast$\\
$A_+$ &  0.25  & 2.08 & 1.14  & $\ast$\\
      &  0.24  & 1.40 & 0.50  & \cite{Melikhov:2001zv}\\\hline
$A_-$ & -0.32  & 2.27 & 1.38  & $\ast$\\
$V$   &  0.32  & 2.21 & 1.30  & $\ast$\\
      &  0.31  & 1.59 & 0.59  & \cite{Melikhov:2001zv}\\\hline
$a_0$ &  0.25  & 1.26 & 0.48  & $\ast$\\
      &  0.27  & 0.74 & 0.19  & \cite{Melikhov:2001zv}\\\hline
$a_+$ &  0.26  & 2.10 & 1.16  & $\ast$\\
$g$   &  0.26  & 2.21 & 1.29  & $\ast$\\ 
      &  0.27  & 1.60 & 0.60  & \cite{Melikhov:2001zv}\\\hline
\end{tabular}
\end{center}
\end{table}
 
\begin{table}[t]
\caption{\label{tab:param_K_Kst} $B\to K(K^\ast)$ transition form factors: values of the parameters in the interpolating function, Eq.\,(\protect\ref{parametrization}).   Our results are marked by an asterisk.   For comparison, where available we also give analogous values inferred from the fit of Ref.\,\cite{Melikhov:2001zv}.}
\begin{center}
\begin{tabular}{lrll|c}
    & $F(0)$  & ~~~$a$ & ~~~$b$ & source \\ \hline
$F_+$ &  0.29 & 1.85  & ~0.96 & $\ast$ \\
      &  0.36 & 1.43  & ~0.43 & \cite{Melikhov:2001zv} \\
$F_-$ & -0.28 & 1.95  & ~1.09 & $\ast$\\
$F_T$ &  0.32 & 1.90  & ~1.02 & $\ast$\\
      &  0.35 & 1.43  & ~0.43 & \cite{Melikhov:2001zv} \\\hline
$A_0$ &  0.40 & 0.98  & 0.034 & $\ast$\\
$A_+$ &  0.30 & 1.92  & ~0.97 & $\ast$\\
      &  0.32 & 1.23  & ~0.38 & \cite{Melikhov:2001zv} \\\hline
$A_-$ & -0.38 & 2.10  & ~1.19 & $\ast$\\
$V$   &  0.37 & 2.05  & ~1.13 & $\ast$\\
      &  0.44 & 1.45  & ~0.45 & \cite{Melikhov:2001zv} \\\hline
$a_0$ &  0.30 & 1.04  & ~0.16 & $\ast$\\
      &  0.39 & 0.72  & ~0.62 & \cite{Melikhov:2001zv} \\\hline
$a_+$ &  0.30 & 1.95  & ~1.00 & $\ast$\\
$g$   &  0.30 & 2.05  & ~1.12 & $\ast$\\ 
      &  0.39 & 1.45  & ~0.45 & \cite{Melikhov:2001zv} \\\hline
\end{tabular}
\end{center}
\end{table}

\subsubsection{Results}
\begin{figure}[t]
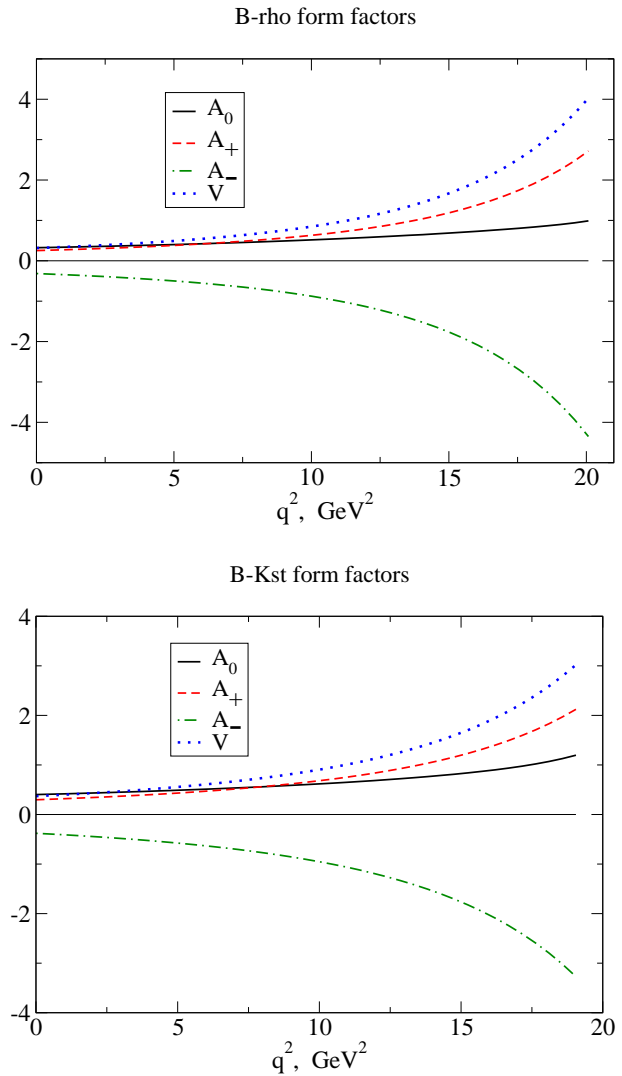

\begin{center}
\hspace*{-0.5cm}
\begin{tabular}{c}
\includegraphics[width=0.45\textwidth]{Fig4aBrho.eps}\\[3ex]

\includegraphics[width=0.45\textwidth]{Fig4bBKst.eps}
\end{tabular}
\end{center}
\caption{\label{fig:formfactors2} Our results for the form factors appearing in 
Eq.\,(\protect\ref{BVV}) -- \emph{Top panel}, $B\to \rho$; and \emph{bottom panel}, $B\to K^\ast$: \emph{solid} $A_0$, \emph{dashed} $A_+$, \emph{dot-dashed} $A_-$, \emph{dotted} $V$.}
\end{figure}

\begin{figure}[t]
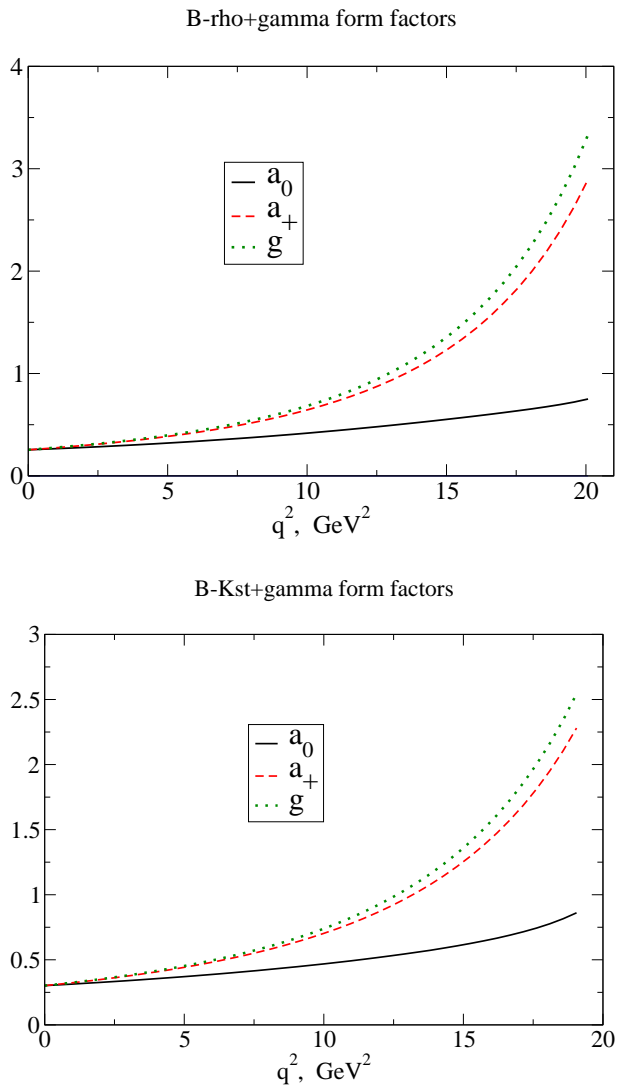

\begin{center}
\hspace*{-0.5cm}
\begin{tabular}{cc}
\includegraphics[width=0.45\textwidth]{Fig5aBrhogam.eps}\\[3ex]

\includegraphics[width=0.45\textwidth]{Fig5bBKstgam.eps}
\end{tabular}
\end{center}
\caption{\label{fig:formfactors3} Our results for the form factors appearing in 
Eq.\,(\protect\ref{BV}) -- \emph{Top panel}, $B\to \rho \gamma$; and \emph{bottom panel}, $B\to K^\ast \gamma$: \emph{solid} $a_0$, \emph{dashed} $a_+$, \emph{dotted} $g$.}
\end{figure}

In Figs.~\ref{fig:formfactors1} -- \ref{fig:formfactors3} we exhibit our calculated form factors for $q^2 \in [0,q^2_{\rm max}]$, with $q^2_{\rm max}= (m_B-m_{P(V)})^2$; viz., on the complete, relevant physical domain.  It is noteworthy and phenomenologically important that in our DSE-based approach all form factors can be calculated on the entire domain of physically accessible momenta.  Moreover, the chiral limit is directly accessible and the consequences of Goldstone's theorem are manifest, so that both pseudoscalar and vector light-quark mesons are realistically described.  No extrapolation in any quantity is required.

Our calculated results are satisfactorily \emph{interpolated} by the simple function
\begin{equation}
\label{parametrization}
F(q^2) = \frac{F(0)}{1-a s +b s^2},\, \; s=q^2/m_B^2.
\end{equation}
We list the values of the form factors at the maximum recoil point, $q^2=0$, and the parameters $a$ and $b$ in Tables~\ref{tab:param_pi_rho} and \ref{tab:param_K_Kst}.  
NB.\ Analytically, 
\begin{equation}
a_0(0)=a_+(0)=g(0)\,,
\end{equation}
a result preserved by the interpolation function.  We provide the interpolations so that our results may readily be adapted as input for other analyses.  

Reference~\cite{Melikhov:2001zv} also provides a fit to these form factors.  While the functional form differs in some instances, one can infer parameter values that are analogous to those in Eq.\,(\ref{parametrization}).  Where this can readily be done, those values are presented for comparison in Tables~\ref{tab:param_pi_rho} and \ref{tab:param_K_Kst}.  As a general rule, in comparison with ours at the maximum recoil point the form factors calculated using the approach reviewed in Ref.\,\cite{Melikhov:2001zv} are larger in magnitude but evolve more slowly.  Large differences can exist between those form factors and ours at $q^2_{\rm max}$.  It is notable that the approach reviewed in Ref.\,\cite{Melikhov:2001zv} may be viewed as describing both light- and heavy-quark propagation via Eq.\,(\ref{mishaprop}).  Resulting partly therefrom, confinement and dynamical chiral symmetry breaking are not veraciously expressed in that approach.

It can be determined from Tables~\ref{tab:param_pi_rho} and \ref{tab:param_K_Kst} that
\begin{equation}
\frac{F_+^{BK}(0)}{F_+^{B\pi}(0)} = 1.23\,,\; 
\frac{A_0^{BK}(0)}{A_0^{B\pi}(0)} = 1.25\,,\; 
\frac{g^{BK^\ast}(0)}{g^{B\rho}(0)} = 1.18\,.
\end{equation}
These ratios are a measure of $SU(3)$-flavour breaking effects in heavy-to-light $B$-decays.  For comparison, the ratio of the constituent-quark masses in Eq.\,(\ref{MQ}) is $1.36$.

In Table~\ref{tab-res} we collect our predictions for the form factors at the maximum recoil point and provide a comparison with extant results obtained within other frameworks.  The figures and tables highlight the wide range of phenomena accessible within our approach.  That is equalled in the table by Refs.\ \cite{Khodjamirian:2006st,Lu:2007sg,Melikhov:2001zv}.  

Table~\ref{relative differences} provides a comparison between our approach and those with equal breadth of application.  Since each study is independent, the degree of quantitative agreement between our results and those of Ref.\,\cite{Lu:2007sg} supports a view that columns two and five in the Table~\ref{tab-res} provide the most reliable predictions.  From this table it appears that exclusive semileptonic $B \rightarrow V$ transitions provide the best means of differentiating between these two frameworks.  On the other hand, it is notable that there is uniformly less agreement between the predictions of the model reviewed in Ref.\,\cite{Melikhov:2001zv}, column six, and those in any of the other approaches.

\section{Epilogue}
\label{sec:end}
We presented a wide-ranging analysis of $B$-meson exclusive semileptonic and rare radiative decays using a phenomenological framework whose elements are based on Dyson-Schwinger equation (DSE) studies in QCD.  Confinement and dynamical chiral symmetry breaking are expressed within this approach.  Moreover, in the present context it is a particular feature of the method that all transition form factors are directly calculable on the entire physical domain of accessible momentum transfer.  This may be contrasted with numerical simulations of lattice regularised QCD, which are currently restricted to the domain of intermediate timelike $q^2$ (e.g., \cite{Becirevic:2006nm} $q^2\in [11,18]\,$GeV$^2$) and QCD Sum Rules, which are directly applicable only on the domain of small timelike $q^2$ (e.g., \cite{Khodjamirian:2006st} $q^2\in [0,10]\,$GeV$^2$).

The results presented herein represent a well-constrained calculation.  Improvement over an earlier study \cite{mishasvy} was made possible by: the appearance of additional data and lattice results in the interim; and technical improvements in our treatment of the loop integrals.  Our results should thus prove valuable in the analysis and correlation of the rapidly accumulating body of information on charmless $B$-decays.  To assist with this, we provided pointwise accurate parametrisations of our calculated transition form factors.

While the foundation of our study is sound, based as it is on reliable results from DSE studies in QCD, it can be improved.  Ideally, one would begin with an interaction kernel and solve directly for every element that appears in a systematic and symmetry preserving truncation of the Schwinger functions contributing to all relevant transitions.  This programme has been realised for numerous processes involving only light mesons; with successes being, e.g., the prediction of the electromagnetic pion form factor \cite{marispion}, the calculation of $K_{\ell 3}$ transition form factors \cite{kl3}, $\pi \pi$ scattering \cite{pipi}, and anomalous processes involving ground and radially excited pseudoscalar mesons \cite{anomaly,radial}.  The programme has begun for heavy-heavy mesons, e.g., Ref.\,\cite{bhagwatmaris,mindgap,krassnigg}.  However, the more difficult problem of an \textit{ab initio} treatment of heavy-light systems remains largely untouched.

\section*{Acknowledgments}
We are grateful for valuable communications with M.\,S.~Bhagwat and I.\,C.~Clo\"et.
M.A.I. acknowledges partial support provided by \emph{Deutsche Forschungsgemeinschaft} grant no.\ 436 RUS 17/65/06, the Heisenberg-Landau Program and the Russian Fund of Basic Research (Grant No.04-02-17370).  This work was also supported in part by: CONICYT (Chile) under grant PBCT/No.285/200; and the United States Department of Energy, Office of Nuclear Physics, contract no.\ DE-AC02-06CH11357.

\begin{widetext}

\begin{table}[tb]
\caption{\label{tab-res} Our calculated values of $B\to \pi,K$ and $B\to \rho,K^*$ form factors at the maximum recoil point compared with the results obtained by other authors.  Based on the widespread application of our approach herein and elsewhere, we estimate that the relative systematic uncertainty in our calculated results is $\sim 15$\%.  In reporting results of Ref.~\cite{Ball:2004ye} we omit an additional uncertainty associated with the kaon's so-called first Gegenbauer moment, which encodes nonperturbative information about the kaon's structure in that framework.}
\begin{centering}
\begin{tabular}{|c|c||c|c|c|c|c|c|}
\hline
&&&&&&\\ 
        & This work & LCSR \cite{Khodjamirian:2006st} & LCSR \cite{Ball:2004ye} &
 LCQM \cite{Lu:2007sg} &  DQM \cite{Melikhov:2001zv} & RQM \cite{Ebert:2006nz} & 
 RCQM \cite{Faessler:2002ut}\\
\hline 
&&&&&&\\[-1,5mm]
 $f^+_{B\pi}(0)$ & 0.24 & 0.25$\pm$0.05 & 0.258$\pm$0.031 & 0.25 & 0.29 & 0.22 & 0.27\\ 
&&&&&&\\[-1,5mm]
\hline
&&&&&&\\[-1,5mm]
 $f^+_{B K}(0)$  & 0.30 & 0.31$\pm$0.04 & 0.331$\pm$0.041 
 & 0.30 & 0.36 &      & 0.36  \\
&&&&&&\\[-1,5mm]
\hline
&&&&&&\\[-1,5mm]
 $f^T_{B\pi}(0)$ & 0.25 & 0.21$\pm$0.04 & 0.253$\pm$0.028 & 0.25 & 0.28 &      &      \\
&&&&&&\\[-1,5mm]
\hline
&&&&&&\\[-1,5mm]
 $f^T_{B K}(0)$  & 0.32 & 0.27$\pm$0.04 & 0.358$\pm$0.037  
  & 0.33 & 0.35 &      & 0.34 \\
&&&&&&\\[-1,5mm]
\hline
&&&&&&\\[-1,5mm]
 $V^{B \rho}(0)$ & 0.31 & 0.32$\pm$0.10 & & 0.30 & 0.31 & 0.30 &     \\
&&&&&&\\[-1,5mm]
\hline
&&&&&&\\[-1,5mm]
 $V^{B K^*}(0)$  & 0.37 & 0.39$\pm$0.11 & & 0.34 & 0.44 &      &    \\
&&&&&&\\[-1,5mm]
\hline
&&&&&&\\[-1,5mm]
 $A_1^{B \rho}(0)$ & 0.24 & 0.24$\pm$0.08 & & 0.23 & 0.26 & 0.27 & \\
&&&&&&\\[-1,5mm]
\hline
&&&&&&\\[-1,5mm]
 $A_1^{B K^*}(0)$  & 0.29 & 0.30$\pm$0.08 & & 0.25 & 0.36 &     & \\
&&&&&&\\[-1,5mm]
\hline
&&&&&&\\[-1,5mm]
 $A_2^{B \rho}(0)$ & 0.25 & 0.21$\pm$0.09 & & 0.22 & 0.24 & 0.28 & \\
&&&&&&\\[-1,5mm]
\hline
&&&&&&\\[-1,5mm]
 $A_2^{B K^*}(0)$  & 0.30 & 0.26$\pm$0.08 & & 0.23 & 0.32 &      & \\
&&&&&&\\[-1,5mm]
\hline
&&&&&&\\[-1,5mm]
 $T_1^{B \rho}(0)$ & 0.26 & 0.28$\pm$0.09 & & 0.26 & 0.27 &     &\\
&&&&&&\\[-1,5mm]
\hline
&&&&&&\\[-1,5mm]
 $T_1^{B K^*}(0)$  & 0.30 & 0.33$\pm$0.10 & & 0.29 & 0.39 & 
$0.24\pm 0.03^{+0.04}_{-0.001}$ \cite{Becirevic:2006nm}    & \\
&&&&&&\\
\hline
\end{tabular}
\end{centering}
\end{table}

\end{widetext}

\begin{table}[t,b]
\caption{\label{relative differences} Mean of the absolute value of the pairwise relative differences between predictions listed in Table~\protect\ref{tab-res} for those studies that report all quantities; viz., 
$|\varepsilon|:= (1/N) \sum_{i=1}^N |1-c^j_i/c^k_i|$, where $j$ and $k$ label columns in the table and $i$ runs over the row number.
Here the present work is denoted by ``$\ast$''.}
\begin{center}
\begin{tabular}{c|cccccc}
comparison & \cite{Khodjamirian:2006st}/$\ast$ & \cite{Lu:2007sg}/$\ast$ &  \cite{Khodjamirian:2006st}/\cite{Lu:2007sg} & \cite{Melikhov:2001zv}/$\ast$ & \cite{Khodjamirian:2006st}/\cite{Melikhov:2001zv} & \cite{Lu:2007sg}/\cite{Melikhov:2001zv}\\\hline
$|\varepsilon|$ (\%)          & 8.2 & 6.3 & 10.2 & 13.2 & 13.7 & 15.1\\
$\sigma_{|\varepsilon|}$ (\%) & 5.8 & 7.0 & ~6.5 & ~9.4 & ~6.8 & ~9.6 \\\hline
\end{tabular}
\end{center}
\end{table}



\begin{thebibliography}{999}
\bibitem{fnEuclidean} NB. In our concrete calculations we use a Euclidean metric, with:  $\{\gamma_\mu,\gamma_\nu\} = 2\delta_{\mu\nu}$; $\gamma_\mu^\dagger = \gamma_\mu$; $a \cdot b = \sum_{i=1}^4 a_i b_i$; and ${\rm tr}[\gamma_5\gamma_\mu\gamma_\nu\gamma_\rho\gamma_\sigma]= 
-4\,\epsilon_{\mu\nu\rho\sigma}\,, \epsilon_{1234}= 1\,.$  For a spacelike vector $k_\mu$, $k^2>0$.
\bibitem{fnrenormalisation} NB. With the models we employ for the mesons' Bethe-Salpeter amplitudes, these integrals are absolutely convergent.  On the other hand, in QCD a renormalisation constant is required to guarantee that these expressions are well defined \cite{mr98,mr97}.
\end{thebibliography}
\end{document}